# Emission redistribution from a quantum dot-bowtie nanoantenna

Armin Regler,[a,b] Konrad Schraml,[a] Anna Lyamkina,[c] Matthias Spiegl,[a] Kai Müller,[a] J. Vuckovic,[b,d] Jonathan J. Finley,[a,e] and Michael Kaniber[a,*]

[a] Walter Schottky Institut and Physik Department, Technische Universität München, Am Coulombwall 4, Garching, Germany, 85748
[b] Institute for Advanced Study, Technische Universität München, Lichtenbergstrasse 2a, Garching, Germany, 85748.
[c] A. V. Rzhanov Institute of Semiconductor Physics SB RAS, Pr. Lavrentieva 13, Novosibirsk, Russia, 630090
[d] E. L. Ginzton Laboratory, Stanford University, Stanford, California, USA, 94305
[e] Nanosystems Initiative Munich, Schellingstraße 4, München, Germany, 85748

**Abstract**. We present a combined experimental and simulation study of a single self-assembled InGaAs quantum dot coupled to a nearby ($\sim 25 nm$) plasmonic antenna. Micro-photoluminescence spectroscopy shows a $\sim 2.4\times$ increase of intensity, which is attributed to spatial far-field redistribution of the emission from the quantum dot-antenna system. Power-dependent studies show similar saturation powers of $2.5 \mu W$ for both coupled and uncoupled quantum dot emission in polarization-resolved measurements. Moreover, time-resolved spectroscopy reveals the absence of Purcell-enhancement of the quantum dot coupled to the antenna as compared to an uncoupled dot, yielding comparable exciton lifetimes of $\tau \sim 0.5 ns$. This observation is supported by numerical simulations, suggesting only minor Purcell-effects of $< 2\times$ for emitter-antenna separations $> 25 nm$. The observed increased emission from a coupled quantum dot-plasmonic antenna system is found to be in good qualitative agreement with numerical simulations and will lead to a better understanding of light-matter-coupling in such novel semiconductor-plasmonic hybrid systems

*E-mail: michael.kaniber@wsi.tum.de

## 1 Introduction

Quantum emitters such as nitrogen-vacancy centers in diamond [1], colloidal nanocrystals [2] or self-assembled semiconductor quantum dots [3] are promising candidates for realizing efficient, scalable and on-chip integrable single photon sources and are inevitable for many applications in quantum information science [4]. In order to control and optimize their light-matter-interaction, they have been coupled to dielectric photonic nanostructures such as microdisks [5], micropillars [6] or photonic crystal nanocavities [7], and many quantum optical key-experiments like Purcell-enhanced spontaneous emission [8], strong coupling [9] and photon blockade [10] have already been demonstrated. Recently, the coupling between individual quantum emitters and plasmonic nanostructures [11] has gained increasing interest in the nano-photonics community due to the great potential of sub-wavelength confinement accompanied by strong electromagnetic field enhancements of surface plasmon polaritons [12]. Plasmonic nanostructures, such as antennas, offer broadband coupling [13] and, thus, avoid spectral mismatch between emitter and photonic mode, whilst still offering extraordinarily high field-enhancements due to their capability to concentrate light to sub-wavelength volumes [14]. The challenge to spatially overlap the emitter with the highly concentrated electromagnetic field has recently been solved for both, dielectric [15] and plasmonic nanostructures [16]. The potential of strong light-matter-interaction in plasmonic nanostructures has already been demonstrated for molecules [17] [18], colloidal nanocrystals [19] [20] and nitrogen vacancy centers [21]. In strong contrast, integration of plasmonic antennas and self-assembled semiconductor quantum dots embedded in a host matrix material is still hardly explored [22][23]. However, those quantum dot-antenna hybrid systems

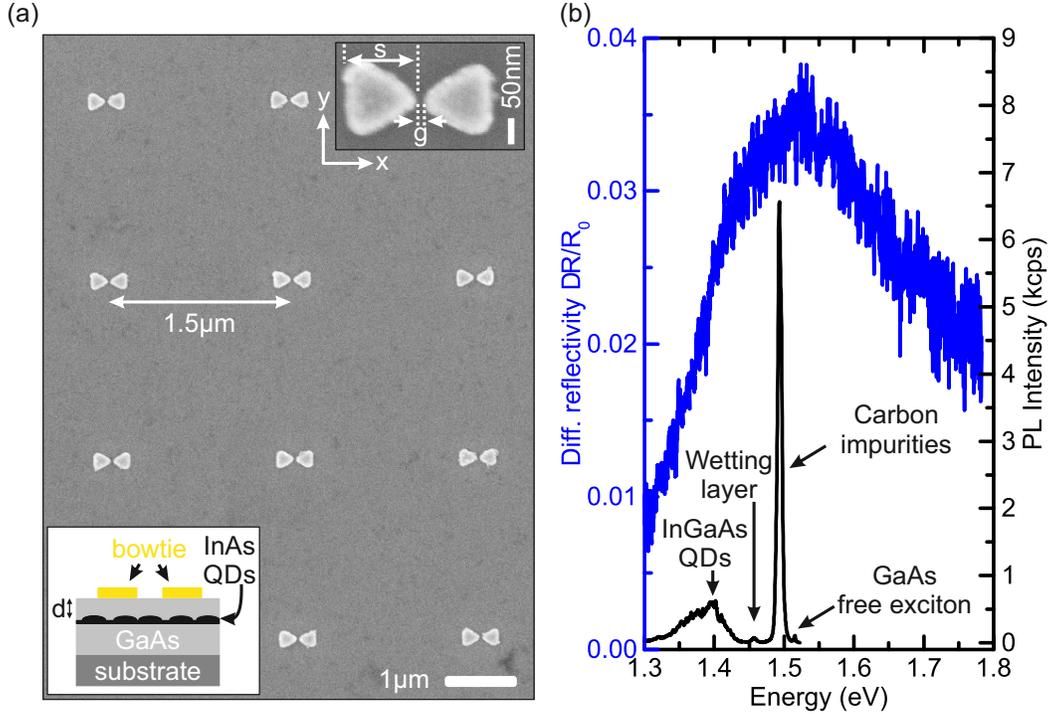

**Figure 1 (a)** Scanning electron microscope image of an array of bowtie nanoantennas with a lattice constant $a = 1.5\mu m$. Left bottom inset: Schematic cross-section of the semiconductor heterostructure and a single bowtie nanoantenna. Right top inset: High-resolution scanning electron microscope image of a single bowtie nanoantennas with triangle size $s = 135nm$ and feed-gap size $g = 12nm$. **(b)** Differential reflectivity spectrum of a bowtie ensemble with nominal triangle size $s_0 = 110nm$ and nominal feed-gap size $g_0 = 30nm$ and photoluminescence spectrum of the InGaAs quantum dot ensemble emission in blue and black, respectively.

would be highly desirable due to their excellent emission properties, their potential for integration in p-i-n-diode structures and the possibility to tune the emission range via band structure engineering to the telecom band, which offers enormous potential for future applications.

Here, we show a 2.4× emission enhancement of an individual InGaAs/GaAs quantum dot spatially located $25nm$ below a single plasmonic bowtie nanoantennas with a feed-gap size $g = 30nm$. Excitation power dependent and polarization-resolved photoluminescence spectroscopy reveal a comparable saturation power of $P_{sat} = 2\mu W$ for both co- and cross-polarized emission of the same quantum dot, indicating that optical absorption enhancement plays only a minor role [22]. Furthermore, we found no significant modification of the spontaneous emission lifetime $\tau$, giving rise to $\tau_c, \tau_{ref} \approx 0.5ns$ for both coupled and reference quantum dots and, thus, ruling out the Purcell-effect as enhancement mechanism. This finding is supported by numerical simulations, demonstrating that only minor (< 2×) enhancements of the spontaneous emission rate are expected for optimally positioned emitters with emitter-antenna separations $d > 25nm$. Finally, we show numerical simulations of the far-field emission profile for an emitter located $25nm$ below a bowtie nanoantenna with $g = 30nm$ and obtain emission enhancements of 18× and 13× for excitation polarization along and perpendicular to the bowtie nanoantenna, respectively. This result strongly supports our attribution of the experimental emission enhancement to spatial redistribution of light due to the bowtie nanoantenna.

## 2   Sample layout, nanofabrication and basic optical properties

The presented semiconductor-nanoplasmonic hybrid system consist of individual semiconductor quantum dots [23] and lithographically defined metallic nanoantennas [24]. The substrate consists of near-surface, optically active InGaAs/GaAs quantum dots grown by solid-source molecular beam epitaxy [25]. On top of a semi-insulating (100) GaAs wafer we deposited a 310nm thick GaAs buffer layer, followed by a single layer of nominally In$_{0.4}$Ga$_{0.6}$As quantum dots, whereby the rotation of the wafer was stopped to obtain a spatially varying quantum dot density [26]. In order to realize three-dimensional electronic confinement, we realized near-surface quantum dots with a $25nm$ thin GaAs capping layer [27]. On top of this epitaxially flat GaAs surface we established state-of-the-art metallic nanoantennas, so-called bowties [28], which consists of two tip-to-tip oriented gold (Au) nanotriangles as shown in the lower inset in Figure 1 (a). The bowtie nanoantennas were fabricated by a combination of electron-beam lithography, electron-beam metallization and lift-off and were arranged in a periodic array with a lattice constant $a = 1.5 \mu m$ as shown in the scanning electron microscope image in Figure 1 (a). A high-resolution zoom of an individual bowtie nanoantennas with triangle size $s = 135nm$ and feed-gap size $g = 12nm$ is shown in the upper inset in Figure 1(a). More details on the nanofabrication process can be found in Ref. [29].

The optical properties of the quantum dot wafer have been studied at cryogenic temperatures ($T = 4K$) using a home-built, dipstick confocal microscope, whereby further details can be found in Ref. [30]. A typical spectrum of the quantum dot ensemble emission, is shown as black curve in Figure 1(b). We observe emission from the GaAs free exciton at $E_{free} = 1.516eV$, Carbon impurities at $E_C = 1.493eV$ [31], the two-dimensional InGaAs wetting layer at $E_{WL} = 1.457eV$ and a broadband emission from the InGaAs quantum dot ensemble at $E_{QD} = 1.385eV$, with an inhomogeneously broadened line width of $\Delta E_{QD} = 70 \pm 4meV$ [26]. The bowtie nanoantennas have been optically characterized using differential white-light reflection spectroscopy for various nominal triangle sizes $s_0$ and feed-gap sizes $g_0$, as defined in the upper inset in Figure 1(a). In Figure 1 (b), we present a typical differential reflectivity spectrum $\Delta R/R_0 \equiv (R_{BT} - R_0)/R_0$ as a function of emission energy plotted as blue curve for $s_0 = 110nm$ and $g_0 = 30nm$. Here, $R_{BT}$ and $R_0$ denote the measured reflectivity signal on the bowtie array and on the bare substrate, respectively. We observe a peak-like response from the bowtie array, indicating a resonant behavior with a resonance energy $E_{res} = 1.520eV$ which is attributed to localized surface plasmon polaritons of the coupled nano-triangles, forming the bowtie nanoantennas [32]. Detailed studies of the optical properties of single bowties and bowtie ensembles can be found in Refs. [29] and [33], respectively. The complementary investigation of micro-photoluminescence from quantum dots and differential reflectivity of bowtie nanoantennas enables us to identify coupled systems with spectrally and spatially matching emission characteristics.

# 3 Optical response from a single quantum dot-antenna system

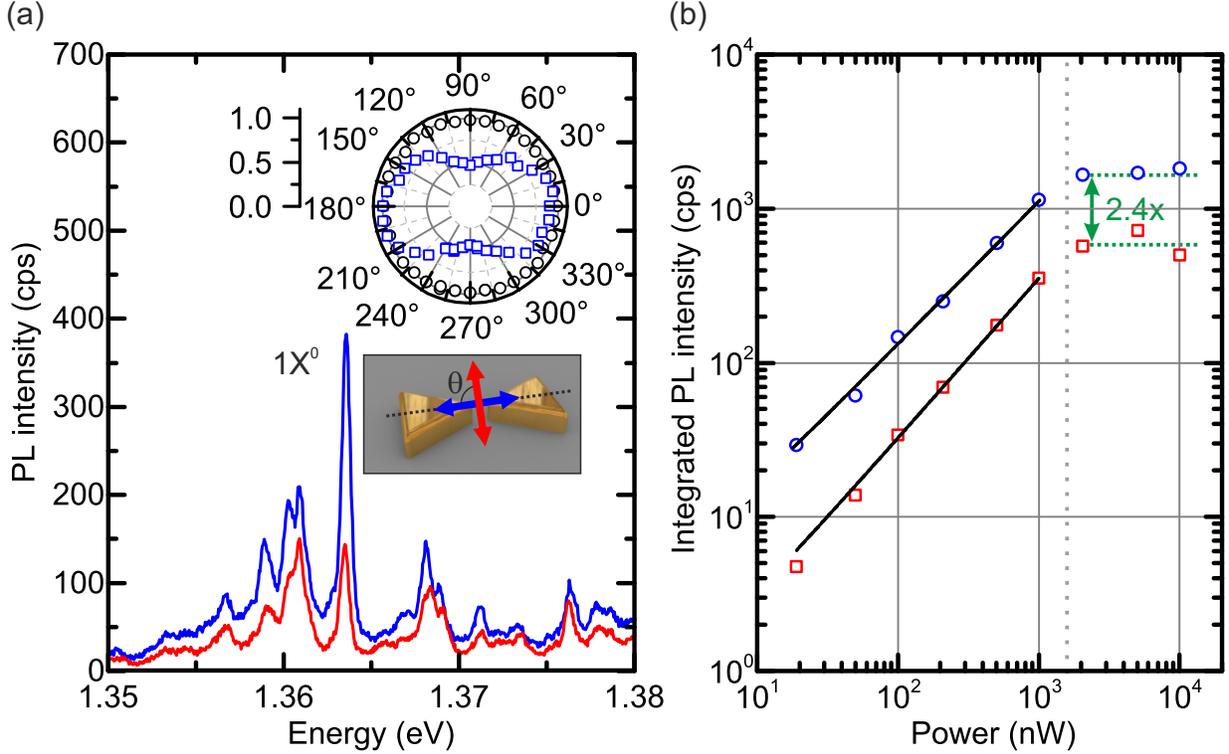

**Figure 2 (a)** Photoluminescence spectra of a single quantum dot spatially positioned below a single bowtie nanoantenna for co- and cross-polarized emission in blue and red, respectively. Inset: Polarization-resolved photluminescence intensity of the $1X^0$-emission and a reference quantum dot in blue and black, respectively. **(b)** Integrated photoluminescence intensity of the emission line labelled $1X^0$ as a function of excitation power for emission polarization parallel ($\theta = 0°$) and perpendicular ($\theta = 90°$) to the bowtie main axis in blue and red, respectively. Black lines correspond to power-law fits.

First we recorded spectrally integrated quantum dot photoluminescence as a function of the spatial position, whilst raster-scanning the excitation spot across the bowtie nanoantenna array for both parallel ($\theta = 0°$) and perpendicular ($\theta = 90°$) linear detection polarization, whilst keeping the excitation polarizer parallel to the bowtie axis (data not shown). A quantum dot coupled to a nanoantenna adapts its emission properties [34] and, thereby, the comparison of co- and cross-polarized emission enabled us to identify emission from individual quantum dots that are interacting with single bowtie nanoantennas. A typical photoluminescence spectra of a single quantum dot-bowtie nanoantenna system is shown in Figure 2(a). Here, we compare the emission of the very same quantum dot for co- ($\theta = 0°$) and cross- ($\theta = 90°$) polarized detection in blue and red, respectively. We observe in both detection geometries emission from sharp emission lines, which are attributed to the discrete (multi-) exciton transitions of individual quantum dots [23]. In particular, we observe that the emission line labelled $1X^0$ ($E = 1.363 eV$) exhibits a pronounced polarization-resolved response as shown by the blue data points in the inset of Figure 2(a) with a degree of linear polarization $\rho_c = \frac{I_{max} - I_{min}}{I_{max} + I_{min}} \sim 39\%$. In strong contrast, all other emission lines exhibit comparable intensities for both co- and cross-polarized detection. Moreover, we present the polarization resolved response of an uncoupled reference quantum dot in the inset of Figure 2(a), clearly demonstrating unpolarized emission with a $\rho_{ref} \sim 3\%$. This finding strongly indicates that the $1X^0$-emission stems from a single quantum dot coupled to the bowtie nanoantenna. In order to gain deeper insights into the nature of the $1X^0$-emission and the according intensity enhancement mechanism, we performed power-dependent photoluminescence spectroscopy. In Figure 2(b), we plot the integrated photoluminescence intensity of the $1X^0$-emission line as a function of the applied excitation power on a double-logarithmic scale for co- and cross-polarized detection in blue and red, respectively. We observe that the $1X^0$-emission line increases linearly for both polarization geometries with slopes $m_{co} = 0.93 \pm 0.01$ and $m_{cross} = 1.04 \pm 0.01$, strongly suggesting a single excitonic character [23]. Moreover, we find that the saturation intensity for co-polarized detection is enhanced by a factor of $\sim 2.4\times$ as compared to the cross-polarized case, as indicated by the green arrow in Figure 2(b). This enhancement is accompanied by the observation of a comparable saturation power of $P_{sat} = 2\mu W$ for both detection polarizations and, therefore, strongly indicates that the observed intensity increase is not related to an enhanced quantum dot excitation rate as recently reported for a similar system in Ref. [22].

**4 Spontaneous emission dynamics of a coupled quantum dot-antenna system**

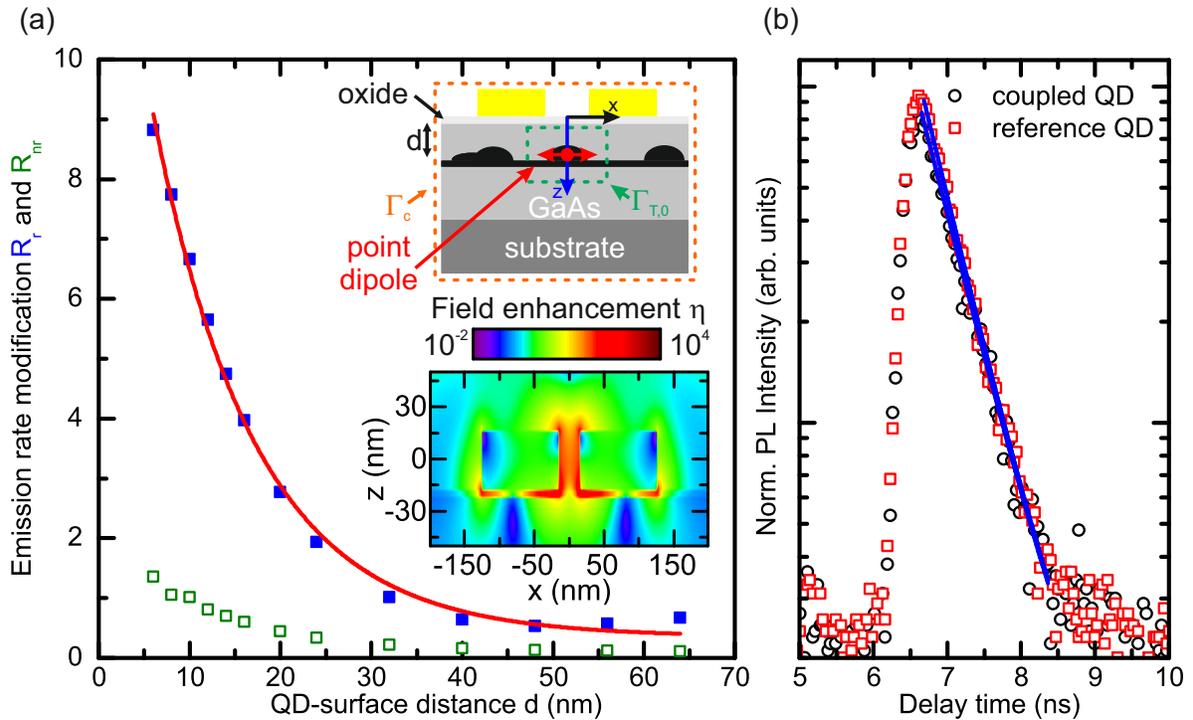

**Figure 3 (a)** Numerical simulation of the spontaneous emission rate modification as a function of quantum dot-surface distance d for radiative and non-radiative processes as blue and green symbols, respectively. Upper inset: Schematic illustration of the simulation geometry. Lower inset: Simulation of the field enhancement $\eta$ in the x-z-plane for y=0. **(b)** Time-resolved photoluminescence spectra for a coupled quantum dot and a reference quantum dot in black and red, respectively. The blue lines show exponential fits to the data.

Another potential mechanism explaining the observed intensity enhancement shown in Figure 2(a) could be an increased spontaneous emission rate due to the Purcell effect [35]. The potential for Purcell enhancement due to the coupling of the near-surface quantum dots to the strongly localized electromagnetic field inside the bowtie feed-gap is investigated by detailed numerical FDTD simulations [36]. For simplicity, we modelled the quantum dot as a point dipole source located at a well-defined distance $d$ below the sample surface as shown in the upper inset of Figure 3(a). Although this point-dipole-approximation yields first indications of the radiative coupling between quantum dot and plasmonic antenna, we note that the mesoscopic size of the quantum dot might lead to a modified light-matter-interaction in strongly confined plasmonic fields [37]. In our simulation the point dipole emits a short pulse of lengths $\Delta t_p = 6.17 fs$, corresponding to energetically broadband emission between $1.03 eV$ and $3.10 eV$. The dipole is surrounded by a box-type monitor (green box in upper inset of Figure 3(a)) that measures the optical power $P_0$ and, thus, the spontaneous emission rate $\Gamma_0 \propto P_0$ of the dipole for reference. A second monitor (orange box) measures the radiative rate of the coupled quantum dot-antenna system $\Gamma_{r,c} \propto P_c$. Consequently, the non-radiative recombination rate $\Gamma_{nr,c}$ is obtained by the difference between both monitors, i.e. $P_0 - P_c \propto \Gamma_{r,0} - \Gamma_{r,c} \equiv \Gamma_{nr,c}$. Therefore, we can define the modification of the radiative and non-radiative emission rate due to the presence of the bowtie nanoantenna with respect to the reference dipole as the ratios $R_r = \Gamma_{r,c}/\Gamma_0$ and $R_{nr} = \Gamma_{nr,c}/\Gamma_0$, respectively. In Figure 3(a), we plot the simulated $R_r$ and $R_{nr}$ as a function of quantum dots-surface separation $d$ and emission energy $E = 1.363 eV$ as blue and green symbols, respectively. First of all, we observe $R_r \sim 1$ for separations $d > 30 nm$ whilst simultaneously $R_{nr} \sim 0$. This indicates that for large separations non-radiative recombination is negligible and the radiative rate stays almost unaffected by the bowtie nanoantenna, close to its reference value. In strong contrast, we observe an increasing $R_r$ for $d \leq 30 nm$, with values reaching $R_r \sim 10\times$ for the smallest separation $d = 6 nm$. The increase of $R_r$ with decreasing $z$ is well described by an exponential increase $R_r(z) = \alpha + \beta e^{-z/z_0}$, as shown by the red line in Figure 3 (a). We obtain for the decay length $z_0 = 11.3 \pm 0.5 nm$, which we attribute to the strong localization of the electromagnetic field at the feed-gap of the bowtie nanoantenna. This hypothesis is supported by complementary simulations of the electromagnetic field enhancement $\eta = |E|^2/|E_0|^2$, where $E$ and $E_0$ denote the field with and without the bowtie, respectively [29]. A typical simulation of $\eta$ for a bowtie nanoantenna with $s_0 = 110 nm$ and $g_0 = 30 nm$ is shown in the lower inset of Figure 3(a). Here, the field is strongly localized within the $4 nm$ thick oxide layer on top of the GaAs substrate, which mimics the natural GaO that grows on top of every quantum dots wafer [38], giving rise to enhancements $\eta \sim 10^3 - 10^4$. Moreover, we observe an increase of the non-radiative rate with decreasing $z \leq 10 nm$ which indicates a quenched luminescence due to enhanced non-radiative energy transfer to the metal of the nanoantenna [39].

In order to experimentally probe if the intensity enhancement of the quantum dot is related to a modification of the spontaneous emission rate, we performed time-resolved photoluminescence spectroscopy using a time correlated single photon counting technique [40]. Typical decay curves for the coupled quantum dot emission $1X^0$ and a reference quantum dot spatially displaced from the bowtie nanoantenna are shown in Figure 3(b) on a semi-logarithmic plot as black circles and red squares, respectively. Both data sets can be fitted by exponential decays as shown by the blue lines in Figure 3(b), yielding comparable spontaneous emission decay times $\tau_c = 0.5 \pm 0.1\ ns$ and $\tau_{ref} = 0.48 \pm 0.1\ ns$ for the coupled and reference quantum dot, respectively. This finding strongly indicates that the intensity enhancement of $1X^0$ is not related to a modified spontaneous emission rate via the Purcell effect. Moreover, our corresponding numerical simulations presented in Figure 3(a) strongly support this finding, indicating only a modest theoretical enhancement of $< 2\times$ for a quantum dot-antenna separation of $d = 25nm$.

## 5 Antenna induced spatial redistribution of emission

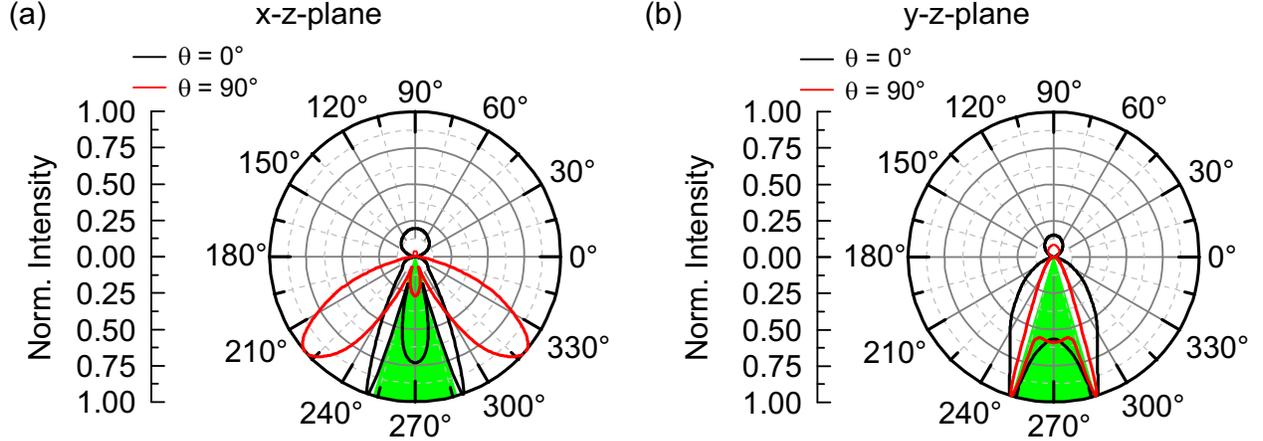

**Figure 4** Numerical simulations of the far-field emission pattern in **(a)** along (x-z-plane) and **(b)** perpendicular (y-z-plane) with respect to the bowtie main axis for a dipole source located $25nm$ below a GaAs-Air interface. Black and red curves show the calculations for excitation polarization parallel ($\boldsymbol{\theta} = \mathbf{0°}$) and perpendicular ($\boldsymbol{\theta} = \mathbf{90°}$) to the bowtie nanoantenna, respectively. The green shaded region indicates the critical angle for total internal reflection at a GaAs-Air interface $\boldsymbol{\theta_c^{GaAs}} \sim \mathbf{16.6°}$.

Finally, we investigate a third potential mechanism of intensity enhancement due to spatial redistribution of the quantum dot emission [19] via the bowtie nanoantenna. Therefore, we performed finite-difference time-domain simulations of the far field emission profile of a point dipole source located $25nm$ below the GaAs-Air interface at the center of the antenna's feed-gap (in the x-y-plane) and polarized along the bowtie main axis. In Figure 4 (a) and (b) we presented the obtained results for excitation polarization parallel ($\theta = 0°$, black curves) and perpendicular ($\theta = 90°$, red curves) to the main axis of a bowtie nanoantenna ($s = 110nm$ and $g = 30nm$) in the x-z and y-z plane, respectively. For better comparison, both curves are normalized to their maximum value. In general, we observe that a significantly larger fraction ($> 70\%$) of the dipole emission is emitted into the high-refractive index GaAs substrate. Moreover, we find for the x-z- and y-z-planes enhancements of the far field emission into the air half space of 18× and 13×, respectively, for excitation polarization parallel to the bowtie nanoantenna (black curves) as compared to the perpendicular excitation polarization (red curve). This strongly indicates that plasmonic antennas result in an enhanced photoluminescence intensity for quantum dot emission polarized along the bowtie main axis. The difference between the emission enhancement observed in experiment and simulation most likely occurs due to a misalignment of the quantum dot position with respect to the feed-gap of the antenna. Moreover, the quantum dots in the experiments have typical lateral dimension of $\sim 20 - 30\ nm$ [41] and, thus, can only be mimicked as a point dipole in first approximation [37]. A deeper insight into the discrepancy in quantitative enhancement between experiment and simulation requires a statistical study of numerous quantum dots as a function of their location with respect to the bowtie feed-gap, geometrical parameters of the antenna and a more accurate theoretical description of the quantum dot as a mesoscopic quantum emitter [42]. In Table 1 we summarize the total emission intensities into air ($I_T^{air}$) and GaAs ($I_T^{GaAs}$) for both planes and excitation polarization directions. The fact that more than 70% of the emission is directed into the GaAs half space indicates large potential for signal enhancement due to micro-patterning of the sample backside, the use of solid immersion lenses, the incorporation of distributed Bragg reflectors or even further engineering of the bowtie nanoantenna regarding directionality and, thus, to enhance the outcoupling of photons towards the upper half space of the sample.

**Table 1** The fraction of emission (in %) into the air and GaAs half space for polarization parallel ($\theta = 0°$) and perpendicular ($\theta = 90°$) to the bowtie nanoantennas into the x-z and y-z-plane.

|  |  | $\theta = 0°$ | | $\theta = 90°$ | |
|---|---|---|---|---|---|
|  |  | x-z-plane | y-z plane | x-z-plane | y-z-plane |
| Air | $I_T^{air}$ (%) | 30.5 | 17.4 | 3.9 | 18.2 |
| GaAs | $I_T^{GaAs}$ (%) | 69.5 | 82.6 | 96.1 | 81.8 |

# 6 Summary, conclusions and outlook

We presented first experimental observations of 2.4 × enhanced emission from near-surface ($\sim 25nm$) self-assembled InGaAs/GaAs quantum dots coupled to a plasmonic bowtie nanoantenna with a feed-gap $g = 30nm$. We employed detailed time-integrated and -resolved micro-photoluminescence measurements in order to probe the enhancement mechanism of this semiconductor-plasmonic hybrid system. We demonstrated that the photoluminescence intensity of the very same quantum dot $1X^0$ as a function of excitation power exhibits a comparable saturation power $P_{sat} = 2\mu W$ for co- and cross-polarized emission, indicating that the observed luminescence enhancement is not due to increased optical excitation absorption [22]. Furthermore, we probed a possible enhancement due to the Purcell-effect [35] via time correlated single photon counting spectroscopy. We found an unmodified spontaneous emission lifetime for the $1X^0$ of $\tau_c = 0.5 \pm 0.1\ ns$ as compared to a reference dot and attribute this observation to a too big separation $d = 25nm$ between quantum dot and sample surface as compared to typical decay lengths of plasmonic fields. This interpretation is supported by numerical simulation which shows only minor theoretical Purcell-enhancements of $< 2\times$ for dot-surface separations $\geq 25nm$. Finally, we performed numerical simulations of the far-field emission of a coupled quantum dot-antenna system which results in pronounced enhancements of 18× and 13× into the air and GaAs half space, respectively, supporting our attribution of the experimental emission enhancement to spatial redistribution of light due to the bowtie nanoantenna.

Our results already emphasis the great potential of buried, self-assembled semiconductor quantum dots [22] [16] [5] in combination with lithographically defined bowtie nanoantennas [28] [29] [33], which offer spectrally broadband and spatially highly confined modes for future plasmonic cavity quantum electrodynamics experiments [43] [44] [3] [11]. In strong contrast to so far mostly used chemically synthesized semiconductor nanocrystals [45], self-assembled InGaAs quantum dots grown by molecular beam epitaxy offer a couple of advantages; (i) They are embedded in a three-dimensional semiconductor matrix and, thus, do not suffer from blinking and bleaching [5], exhibiting already short exciton lifetimes on the order of $1ns$ and internal quantum efficiencies close to unity at cryogenic temperatures [46]. (ii) Moreover, since the quantum dots are directly grown on a semiconductor substrate, such as GaAs [46] and even Si [47], they are structurally stable and one can easily integrate them into photonic [3] and plasmonic nanostructures [29] [33]. (iii) In addition to their spectrally broadband response [28], plasmonic cavities are intrinsically capable to exploit the employed metal also as local electrical contacts [48] and, thus, allows for electrical tunability [41]. As a consequence, we envision a broad scope of future experiments for this semiconductor-plasmonic hybrid system, studying for example its resonant response using resonant fluorescence spectroscopy [49] and, thus, evaluating the potential for coherent control experiments [50]. Furthermore, it has recently been demonstrated that the mesoscopic size of self-assembled quantum dots in combination with the strongly confined plasmonic fields leads to a break-down of the commonly applied simplistic point-dipole approximation [37], probably giving rise to modified optical selection rules and, thus, a novel regime of light-matter-interaction.

*Acknowledgments*

We acknowledge financial support of the DFG via the SFB 631, Teilprojekt B3, the German Excellence Initiative via NIM, and FP-7 of the European Union via SOLID. The authors gratefully acknowledge support of the Technische Universität München (TUM) − Institute for Advanced Study, funded by the German Excellence Initiative and the TUM International Graduate School of Science and Engineering (IGSSE).

*References*

[1] F. Jelezko and J. Wrachtrup, "Single defect centres in diamond: A review," *Phys. Stat. Sol. (a),* vol. 203, p. 3207, 2006.

[2] J. Y. Kim, O. Voznyy, D. Zhitomirsky and E. H. Sargent, "25th Anniversary Article: Colloidal Quantum Dot Materials and Devices: A Quarter-Century of Advances," *Adv. Mater.,* vol. 25, p. 4986, 2013.

[3] P. Lodahl, S. Mahmoodian and S. Stobbe, "Interfacing single photons and single quantum dots with photonic nanostructures," *Rev. Mod. Phys.,* vol. 87, p. 347, 2015.

[4] J. L. O'Brien, A. Furusawa and J. Vuckovic, "Photonic quantum technologies," *Nature Photonics,* vol. 3, p. 687, 2009.

[5] P. Michler, A. Kirar, C. Becher, W. V. Schoenfeld, P. M. Petroff, L. Zhang, E. Hu and A. Imamoglu, "A Quantum Dot Single-Photon Turnstile Device," *Science,* vol. 290, p. 2282, 2000.

[6] M. Pelton, C. Santori, J. Vuckovic, B. Zhang, G. S. Solomon, J. Plant and Y. Yamamoto, "Efficient Source of Single Photons: A Single Quantum Dot in a Micropost Microcavity," *Phys. Rev. Lett.,* vol. 89, p. 233602, 2002.

[7] D. Englund, D. Fattal, E. Waks, G. Solomon, B. Zhang, T. Nakaoka, Y. Arakawa, Y. Yamanoto and J. Vuckovic, "Controlling the Spontaneous Emission Rate of Single Quantum Dots in a Two-dimensional Photon Crystal," *Phys. Rev. Lett.,* vol. 95, p. 013904, 2005.

[8] A. Kress, F. Hofbauer, N. Reinelt, M. Kaniber, H. J. Krenner, R. Meyer, G. Böhm and J. J. Finley, "Manipulation of the spontaneous emission dynamics of quantum dots in two-dimensional photonic crystals," *Phys. Rev. B,* vol. 71, p. 241304(R), 2005.

[9] T. Yoshie, A. Scherer, J. Hendrickson, G. Khitrova, H. M. Gibbs, G. Rupper, C. Ell, O. B. Shchekin and D. G. Deppe, "Vacuum Rabi splitting with a single quantum dot in a photonic crystal nanocavity," *Nature,* vol. 432, p. 200, 2004.

[10] A. Faraon, I. Fushman, D. Englund, N. Stoltz, P. Petroff and V. Vuckovic, "Coherent generation of non-classical light on a chip via photon-induced tunnelling and blockade," *Nature Physics,* vol. 4, p. 859, 2008.

[11] M. S. Tame, K. R. McEnery, S. K. Özdemir, J. Lee, S. A. Maier and M. S. Kim, "Quantum Plasmonics," *Nature Physics,* vol. 9, p. 329, 2013.

[12] J. A. Schuller, E. S. Barnard, W. Cai, Y. C. Jun, J. S. White and M. L. Brongersma, "Plasmonics for extreme light concentration and manipulation," *Nature Materials,* vol. 9, p. 193, 2010.

[13] D. P. Fromm, A. Sundaramurthy, P. J. Schuck, G. Kino and W. E. Moerner, "Gap-Dependent Optical Coupling of Single "Bowtie" Nanoantennas Resonant in the Visible," *Nano Letters,* vol. 4, p. 957, 2004.

[14] A. Sundaramurthy, K. B. Crozier, G. S. Kino, D. P. Fromm, P. J. Schuck and W. E. Moerner, "Field enhancement and gap-dependent resonance in a system of two opposing tip-to-tip Au nanotriangles," *Phys. Rev. B,* vol. 72, p. 165409, 2005.

[15] A. Badolator, K. Hennessy, M. Atatüre, J. Dreiser, E. Hu, P. M. Petroff and A. Imamoglu, "Deterministic Coupling of Single Quantum Dots to Single Nanocavity Modes," *Science,* vol. 308, p. 1158, 2005.

[16] M. Pfeiffer, K. Lindfors, H. Zhang, B. Fenk, F. Phillipp, P. Atkinson, A. Rastielli, O. G. Schmidth, H. Giessen and M. Lippitz, "Eleven Nanometer Alignment Precision of a Plasmonic Nanoantenna with a Self-Assembled GaAs Quantum Dot," *Nano Lett. ,* vol. 14, p. 197, 2014.

[17] A. Kinkhawala, Z. Yu, S. Fan, Y. Avlasevich, K. Müllen and W. E. Moerner, "Large single-bowtie fluorescence enhancements produced by a bowtie nanoantenna," *Nature Photonics,* vol. 3, p. 654, 2009.

[18] G. M. Akselrod, C. Argyropoulos, T. B. Hoang, C. Ciraci, C. Fang, J. Huang, D. R. Smith and M. H. Mikkelsen, "Probing the mechanisms of large Purcell enhancement in plasmonic nanoantennas," *Nature Photonics,* vol. 8, p. 835, 2014.

[19] A. G. Curto, G. Volpe, T. H. Taminiau, M. P. Kreuzer, R. Quidant and N. F. van Hulst, "Unidirectional Emission of a Quantum Dot Coupled to a Nanoantenna," *Science,* vol. 329, p. 930, 2010.

[20] T. B. Hoang, G. M. Akselrod, C. Argyropoulos, J. Huang, D. R. Smith and M. H. Mikkelsen, "Ultrafast spontaneous emission source using plasmonic nanoantennas," *Nature Communications,* vol. 6, p. 7788, 2015.

[21] A. W. Schell, G. Kewes, T. Hanke, F. Leitenstorfer, R. Bratschitsch, O. Benson and T. Aichele, "Single defect centers in diamond nanocrystals as quantum probes for plasmonic nanostructures," *Optics Express,* vol. 19, p. 7914, 2011.

[22] M. Pfeiffer, K. Lindfors, C. Wolpert, P. Atkinson, M. Benyoucef, A. Rastelli, O. G. Schmidt, H. Giessen and M. Lippitz, "Enhancing the Optical Exictation Efficiency of a Single Self-Assembled Quantum Dot with a Plasmonic Nanoantenna," *Nano Letters,* vol. 10, p. 4555, 2010.

[23] J. J. Finley, A. D. Ashmore, A. Lemaître, D. J. Mowbray, M. S. Skolnick, I. E. Itskevich, P. A. Maksym, M. Hopkinson and T. F. Krauss, "Charged and neutral exciton complexes in individual self-assembled In(Ga)As quantum dots," *Phys. Rev. B,* vol. 63, p. 073307, 2001.

[24] P. Biagioni, J.-S. Huang and B. Hecht, "Nanoantennas for visible and infrared radiation," *Rep. Prog. Phys.,* vol. 75, p. 024402, 2012.

[25] K. Y. Cheng, "Development of molecular beam epitaxy technology for III-V compound semicondcutor heterostructur devices," *J. Vac. Sci. Technol. A,* vol. 31, p. 050814, 2013.

[26] H. J. Krenner, S. Stufler, M. Sabathil, E. C. Clark, P. Ester, M. Bicherl, G. Abstreiter, J. J. Finley and A. Zrenner, "Recent advances in exciton-based quantum information processing in quantum dot nanostructures," *New J. Phys.,* vol. 7, p. 184, 2005.

[27] K. Adlkofer, E. F. Duijs, F. Findeis, M. Bichler, M. Zrenner, E. Sackmann, G. Abstreiter and M. Tanaka, "Enhancement of photoluminescence from near-surface quantum dots by suppression of surface state density," *Phys. Chem. Chem. Phys.,* vol. 4, p. 785, 2002.

[28] D. P. Fromm, A. Sundaramurthy, P. J. Schuck, G. Kino and W. E. Moerner, "Gap-Dependent Optical Coupling of Single "Bowtie" Nanoantennas Resonant in the Visible," *Nano Letters,* vol. 5, p. 957, 2004.

[29] K. Schraml, M. Spiegl, M. Kammerlocher, G. Bracher, J. Bartl, T. Campbell, J. J. Finley and M. Kaniber, "Optical properties and interparticle coupling of plasmonic bowtie nanoantennas on a semiconducting substrate," *Phys. Rev. B,* vol. 90, p. 035435, 2014.

[30] F. Flassig, M. Kaniber, G. Reithmaier, K. Müller, A. Andrejew, R. Gross, J. Vuckovic and J. J. Finley, "Towards on-chip generation, routing and detection of non-classical light," *Proc. of SPIE,* vol. 9373, p. 937305, 2015.

[31] T. Mano, K. Watanabe, S. Tsukamoto, H. Fujioka, M. Oshima and N. Koguchi, "Fabrication of InGaAs quantum dots on GaAs(001) by droplet epitaxy," *J. Cryst. Growth,* vol. 209, p. 504, 2000.

[32] P. Nordlander, C. Oubre, E. Prodan, K. Li and M. I. Stockman, "Plamon Hybridization in Nanoparticle Dimers," *Nano Letters,* vol. 4, p. 899, 2004.

[33] M. Kaniber, K. Schraml, A. Regler, J. Bartl, G. Glashagen, F. Flassig, J. Wierzbowski and J. Finley, "Surface plasmon resonance spectroscopy of single bowtie nano-antennas using a differential reflectivity method," *submitted,* 2015.

[34] T. H. Taminiau, F. D. Stefani, F. B. Segerink and N. F. Van Hulst, "Optical antennas direct single-molecule emission," *Nature Photonics,* vol. 2, p. 234, 2008.

[35] E. M. Purcell, "Spontaneous emission probabilities at radio frequencies," *Phys. Rev.,* vol. 69, p. 681, 1946.

[36] Lumerical Solutions, Inc., "http://www.lumerical.com/tcad-products/fdtd/," [Online].

[37] M. L. Andersen, S. Stobbe, A. S. Sorensen and P. Lodahl, "Strongly modified plasmon-matter interaction with mesoscopic quantum emitters," *Nature Physics,* vol. 7, p. 215, 2011.

[38] D. G. Davies, D. M. Whittaker and L. R. Wilson, "Metal nanoantenna plasmon resonance lineshape modification by semiconductor surface native oxide," *J. Appl. Phys.,* vol. 112, p. 044315, 2012.

[39] P. Anger, P. Bharadwaj and L. Novotny, "Enhancement and Quenching of Single-Molecule Fluorescence," *Phys. Rev. Lett.,* vol. 96, p. 113002, 2006.

[40] D. V. O'Connor and D. Phillips, Time-Correlated Single Photon Counting, London/Orlando: Academic Press, 1984.

[41] M. Kaniber, M. F. Huck, K. Müller, E. C. Clark, F. Troiani, M. Bichler, H. J. Krenner and J. J. Finley, "Electrical control of the exciton-biexciton splitting in self-assembled InGaAs quantum dots," *Nanotechnology,* vol. 22, p. 325202, 2011.

[42] P. Tighineanu, A. S. Sorensen, S. Stobbe and P. Lodahl, "Unraveling the Mesoscopic Character of Quantum Dots in Nanophotonics," *Phys. Rev. Lett.,* vol. 114, p. 247401, 2015.

[43] G. Khitrova, H. M. Gibbs, M. Kira, S. W. Koch and A. Scherer, "Vacuum Rabi splitting in semiconductors," *Nature Physics,* vol. 2, p. 81, 2006.

[44] S. Noda, M. Fujita and T. Asano, "Spontaneous-emission control by photonic crystals and nanocavities," *Nature Photonics,* vol. 1, p. 449, 2007.

[45] P. Michler, A. Imamoglu, M. D. Mason, P. J. Carson, G. F. Strouse and S. K. Buratto, "Quantum correlation among photons from a single quantum dot at room temperature," *Nature,* vol. 406, p. 968, 2000.

[46] J. M. Gérard, O. Cabrol and B. Sermage, "InAs quantum boxes: Highly efficient radiative traps for light emitting devices on Si," *Appl. Phys. Lett.,* vol. 68, p. 3123, 1996.

[47] I. J. Luxmoore, R. Toro, O. Del Pozo-Zamudio, N. A. Wasley, E. A. Chekhovich, A. M. Sanchez, R. Beanland, A. M. Fox, M. S. Skolnick, H. Y. Liu and A. I. Tartakovskii, "III-V quantum light source and cavity-QED on Silicon," *Scientific Reports,* vol. 3, p. 1239, 2013.

[48] J. C. Prangsma, J. Kren, A. G. Knapp, S. Grossmann, M. Emmerling, M. Kamp and B. Hect, "Electrically Connected Resonant Optical Antennas," *Nano Letters,* vol. 12, p. 3915, 2012.

[49] E. B. Flagg, A. Muller, J. W. Robertson, S. Founta, D. G. Deppe, M. Xiao, W. Ma, G. J. Salamo and C. K. Shih, "Resonantly driven coherent oscillations in a solid-state quantum emitter," *Nature Physics,* vol. 5, p. 203, 2009.

[50] M. Aeschlimann, M. Bauer, D. Bayer, T. Brixner, F. J. García de Abajo, W. Pfeiffer, M. Rohmer, C. Spindler and F. Steeb, "Adaptive subwavelength control of nano-optical fields," *Nature,* vol. 446, p. 301, 2007.

[51] G. Bracher, K. Schraml, M. Blauth, J. Wierzbowski, N. Coca López, M. Bicherl, K. Müller, J. J. Finley and M. Kaniber, "Imaging surface plasmon polaritons using proximal self-assembled InGaAs quantum dots," *J. Appl. Phys.,* vol. 116, p. 033101, 2014.